\documentclass[superscriptaddress,aps,preprintnumbers,amsmath,showpacs,amssymb,prd,nofootinbib,reprint]{revtex4-1}
\usepackage{bm, color} 
\usepackage{amssymb,amsfonts,slashed,amsthm,amsmath,graphicx, soul}
\usepackage[caption=false]{subfig}
\usepackage{hyperref}

\begin{document}


\newcommand{\vev}[1]{ \left\langle {#1} \right\rangle }
\newcommand{\bra}[1]{ \langle {#1} | }
\newcommand{\ket}[1]{ | {#1} \rangle }
\newcommand{\eV}{ \ {\rm eV} }
\newcommand{\KeV}{ \ {\rm keV} }
\newcommand{\MeV}{\  {\rm MeV} }
\newcommand{\GeV}{\  {\rm GeV} }
\newcommand{\TeV}{\  {\rm TeV} }
\newcommand{\1}{\mbox{1}\hspace{-0.25em}\mbox{l}}
\newcommand{\Red}[1]{{\color{red} {#1}}}

\newcommand{\lmk}{\left(}  
\newcommand{\rmk}{\right)}
\newcommand{\lkk}{\left[}  
\newcommand{\rkk}{\right]}
\newcommand{\lhk}{\left \{ }  
\newcommand{\rhk}{\right \} }
\newcommand{\del}{\partial}  
\newcommand{\la}{\left\langle} 
\newcommand{\ra}{\right\rangle}
\newcommand{\half}{\frac{1}{2}}

\newcommand{\bea}{\begin{array}}
\newcommand{\eea}{\end{array}}
\newcommand{\beq}{\begin{eqnarray}}
\newcommand{\eeq}{\end{eqnarray}}
\newcommand{\eq}[1]{Eq.~(\ref{#1})}

\newcommand{\dd}{\mathrm{d}}
\newcommand{\Mpl}{M_{\rm Pl}}
\newcommand{\mg}{m_{3/2}}
\newcommand{\abs}[1]{\left\vert {#1} \right\vert}
\newcommand{\mphi}{m_{\phi}}
\newcommand{\Hz}{\ {\rm Hz}}
\newcommand{\for}{\quad \text{for }}
\newcommand{\Min}{\text{Min}}
\newcommand{\Max}{\text{Max}}
\newcommand{\Kahler}{K\"{a}hler }
\newcommand{\cphi}{\varphi}
\newcommand{\Tr}{\text{Tr}}
\newcommand{\diag}{{\rm diag}}

\newcommand{\SUf}{SU(3)_{\rm f}}
\newcommand{\Upq}{U(1)_{\rm PQ}}
\newcommand{\Zpq}{Z^{\rm PQ}_3}
\newcommand{\Cpq}{C_{\rm PQ}}
\newcommand{\ubar}{u^c}
\newcommand{\dbar}{d^c}
\newcommand{\ebar}{e^c}
\newcommand{\nubar}{\nu^c}
\newcommand{\Ndw}{N_{\rm DW}}
\newcommand{\Fpq}{F_{\rm PQ}}
\newcommand{\fpq}{v_{\rm PQ}}
\newcommand{\Br}{{\rm Br}}
\newcommand{\Lag}{\mathcal{L}}
\newcommand{\Lqcd}{\Lambda_{\rm QCD}}

\newcommand{\ji}{j_{\rm inf}} 
\newcommand{\jb}{j_{B-L}} 
\newcommand{\M}{M} 
\newcommand{\im}{{\rm Im} }
\newcommand{\re}{{\rm Re} }

\def\lrf#1#2{ \left(\frac{#1}{#2}\right)}
\def\lrfp#1#2#3{ \left(\frac{#1}{#2} \right)^{#3}}
\def\lrp#1#2{\left( #1 \right)^{#2}}
\def\REF#1{Ref.~\cite{#1}}
\def\SEC#1{Sec.~\ref{#1}}
\def\FIG#1{Fig.~\ref{#1}}
\def\EQ#1{Eq.~(\ref{#1})}
\def\EQS#1{Eqs.~(\ref{#1})}
\def\TEV#1{10^{#1}{\rm\,TeV}}
\def\GEV#1{10^{#1}{\rm\,GeV}}
\def\MEV#1{10^{#1}{\rm\,MeV}}
\def\KEV#1{10^{#1}{\rm\,keV}}
\def\blue#1{\textcolor{blue}{#1}}
\def\red#1{\textcolor{red}{#1}}

\newcommand{\eff}{\Delta N_{\rm eff}}
\newcommand{\neff}{\Delta N_{\rm eff}}
\newcommand{\cc}{\Lambda}
\newcommand{\Mpc}{\ {\rm Mpc}}
\newcommand{\Msolar}{M_\cdot}

\newcommand{\mav}{\left. m_a^2 \right\vert_{T=0}}
\newcommand{\mat}{m_{a, {\rm QCD}}^2 (T)}
\newcommand{\mam}{m_{a, {\rm M}}^2 }


\preprint{TU-1104, IPMU20-0069}

\title{
XENON1T excess from anomaly-free ALP dark matter \\ and its implications for stellar cooling anomaly
}

\author{
Fuminobu Takahashi
}
\affiliation{Department of Physics, Tohoku University, 
Sendai, Miyagi 980-8578, Japan} 
\affiliation{Kavli IPMU (WPI), UTIAS, 
The University of Tokyo, 
Kashiwa, Chiba 277-8583, Japan}

\author{
Masaki Yamada
}
\affiliation{Frontier Research Institute for Interdisciplinary Sciences, Tohoku University, Sendai, Miyagi 980-8578, Japan}
\affiliation{Department of Physics, Tohoku University, 
Sendai, Miyagi 980-8578, Japan}

\author{
Wen Yin
}
\affiliation{Department of Physics, Faculty of Science, The University of Tokyo,  Bunkyo-ku, Tokyo 113-0033, Japan}

\begin{abstract}
Recently, an anomalous excess was found in the electronic recoil data collected at the XENON1T experiment. The excess may be explained by an axion-like particle (ALP) with mass of a few keV and a coupling to electron  of $g_{ae} \sim 10^{-13}$, if the ALP constitutes all or some fraction of local dark matter (DM). In order to satisfy the X-ray constraint, the ALP coupling to photons must be significantly suppressed compared to that to electrons. This strongly suggests that the ALP has no anomalous couplings to photons, i.e., there is no U(1)$_{\rm PQ}$-U(1)$_{\rm em}$-U(1)$_{\rm em}$ anomaly. We show that such anomaly-free ALP DM predicts an X-ray line signal with a definite strength through the operator arising from threshold corrections, and compare it with the projected sensitivity of the ATHENA X-ray observatory. The abundance of ALP DM can be explained by the misalignment mechanism, or by thermal production if it constitutes a part of DM. In particular, we find that the anomalous excess reported by the XENON1T experiment as well as the stellar cooling anomalies from white dwarfs and red giants can be  explained simultaneously better when the ALP constitutes about 10\% of DM. As concrete models, we revisit the leptophilic anomaly-free ALP DM considered in Ref.\cite{Nakayama:2014cza} as well as an ALP model based on a two Higgs doublet model in the supplemental material.
\end{abstract}

\maketitle
\flushbottom

{\bf Introduction.--}
An excess of electron recoil events over the background has been recently reported by 
the XENON1T experiment~\cite{Aprile:2020tmw}. The statistical significance of the excess
is about $3\sigma$, whose precise value depends on the potential source.
The observed excess can be nicely fitted by the solar axion, which, however, is
in strong tension with the star cooling constraints~\cite{Viaux:2013lha,Bertolami:2014wua,Corsico:2016okh,Battich:2016htm,Giannotti:2015kwo}. 
While the significance is weaker than the solar axion, 
the excess may be explained by an axion-like particle (ALP) with mass of a few keV and
the coupling to electron $g_{ae} \sim 10^{-13}$, if it constitutes all or some fraction of 
the local dark matter (DM). Such value of $g_{ae}$ is intriguingly close to the one suggested by the so-called 
 stellar cooling anomalies from white dwarfs and red giants.
In this letter we focus on this possibility of ALP DM responsible for the excess observed by the XENON1T
experiment, and study its implications for cosmology and astrophysics, especially the  stellar
cooling anomaly~\cite{Giannotti:2017hny}.

The ALP is a pseudo-Nambu-Goldstone boson that appears when a continuous global symmetry is
spontaneously broken. Hereafter we call the symmetry the Peccei-Quinn (PQ) symmetry~\cite{Peccei:1977hh,Peccei:1977ur}, although the ALP considered in this letter is not
the QCD axion. See Refs.~\cite{Jaeckel:2010ni,Ringwald:2012hr,Arias:2012az,Graham:2015ouw,Marsh:2015xka, Irastorza:2018dyq, DiLuzio:2020wdo} for reviews.
There are two relevant parameters that characterize the ALP; one is the mass, $m_a$, and
the other is the decay constant, $f_a$. 
The decay constant is of order the symmetry breaking scale unless
the sector responsible for the spontaneous symmetry breaking is rather contrived
as in the clockwork/aligned axion model~\cite{Choi:2015fiu,Kaplan:2015fuy,Higaki:2015jag,Higaki:2016yqk,Giudice:2016yja,Farina:2016tgd}.
Since the mass is  suppressed thanks to the PQ symmetry, the ALP can be naturally light. 
If the lifetime of the ALP
is much longer than the age of the Universe, it can be all or a part of DM.

The ALP is considered to have various couplings to the standard model (SM) particles, and among them,
we focus on the couplings to photons and electrons, but we also give a  comment on
a coupling to nucleons later. Throughout this letter we 
assume the ALP mass and coupling to electron suggested by the XENON1T excess.  Then, the preferred range of $g_{ae}$
implies that
the typical value of the decay constant is $f_a \sim 10^{9-10}$\,GeV.

Let us first assume that the ALP has an anomaly of U(1)$_{\rm PQ}$-U(1)$_{\rm em}$-U(1)$_{\rm em}$, where U(1)$_{\rm em}$ is the electromagnetic gauge symmetry. Then the ALP also couples to photon with a coupling of $g_{a \gamma \gamma} \sim \alpha_{\rm em}/2 \pi f_a$, 
where $\alpha_{\rm em}$ is a fine-structure constant of the electromagnetic interaction. 
Here we omit a model-dependent constant which is of order unity in most cases. This leads to the decay of ALP into two photons in the late universe, which may leave a detectable X-ray line signal. 
In fact,  such ALP DM with an anomalous coupling to photons
is excluded by  the X-ray constraint on the coupling $g_{a\gamma \gamma}$ for $m_a \gtrsim 0.1 \KeV$ (see e.g. Ref.~\cite{Irastorza:2018dyq} and references therein).\footnote{The X-ray bound can be converted to 
$g_{ae} \lesssim {\cal O}(10^{-18})$ at $m_a$ of a few keV, using Eq.~(\ref{conversion}).}
Therefore, we are led to consider ALP DM whose coupling to photons is significantly suppressed
compared to that to electron.

Another interesting observation is that the axion-electron coupling $g_{ae} \sim 10^{-13}$ suggested
by the XENON1T excess has an overlap with that hinted by various cooling excesses found in e.g. white dwarfs (WD)
and red-giants (RG)~\cite{Giannotti:2017hny}. Considering that the stellar cooling argument does not depend on the
cosmological abundance of the ALP DM, and that the ALP mass is usually set to be much lighter than the typical energy scale in the stellar environment, we will see how the overlap becomes even 
better by varying 
the fraction of ALP DM and the ALP mass.

In this letter, we study the ALP DM as a potential source for the XENON1T excess, and discuss
its implications for the stellar cooling anomaly.
To satisfy the
observational constraint, 
the ALP coupling to photons is significantly suppressed compared to that to electrons. This is 
realized if the PQ symmetry is free from the U(1)$_{\rm PQ}$-U(1)$_{\rm em}$-U(1)$_{\rm em}$ anomaly~\cite{Pospelov:2008jk,Nakayama:2014cza}.
As we will see, even in the anomaly-free ALP models,  the ALP does have a nonzero coupling
to photons, which arises from threshold corrections. 
The corresponding operator is suppressed by the mass ratio between the ALP and the charged fermions to which
the ALP is coupled, and thus, it is generically dominated by the contribution of electron. Therefore, the decay
rate for the ALP into two photons is rather robust and universal for a generic anomaly-free ALP model.
We will compare the predicted flux of the X-ray from the ALP decay with the reach of the future  ATHENA X-ray observatory~\cite{Neronov:2015kca}.
We also discuss the viable production mechanisms of the ALP as well as isocurvature constraint on the inflation scale.

In the supplemental material, we revisit the leptophilic ALP DM model as well as an ALP model based on a two Higgs doublet model 
as concrete examples of anomaly-free ALP DM.

\vspace{0.2cm}

{\bf ALP couplings to SM particles.--}
Following Ref.~\cite{Nakayama:2014cza}, 
we consider the case in which there is no U(1)$_{\rm PQ}$-U(1)$_{\rm em}$-U(1)$_{\rm em}$ anomaly. An explicit construction of this kind of models is discussed in the supplemental material. 

We start from the following interactions, 
\beq
 - \mathcal{L}_L = \sum_f m_f \bar{f}_R f_L e^{i q_f \frac{a}{f_a}} \, , 
\eeq
where $a$ is the ALP, and $q_f$ ($f = e, \mu, \tau, u, d, c, s, t, b$) are the PQ charges of the SM particle $f$.\footnote{Here we focus on the ALP coupling to charged SM fermions. If the ALP has a similar coupling to
neutrinos, the ALP will mainly decay into neutrinos.}
We choose the PQ charges so that there is no U(1)$_{\rm PQ}$-U(1)$_{\rm em}$-U(1)$_{\rm em}$ anomaly.\footnote{
If the axion is coupled to quarks, the color anomaly should also vanish (or should be suppressed by several
orders of magnitude), since otherwise the mixing with $\pi^0$ would
induce the axion-photon coupling. In this sense, the ALP considered in our scenario is complementary to the QCD axion.
}
The coupling between the ALP and electron is given by $g_{ae} = q_e m_e/f_a$, 
in other words,
\beq
\label{conversion}
 \frac{f_a}{q_e} \simeq 10^{10} \GeV \lmk \frac{g_{ae}}{5 \times 10^{-14}} \rmk^{-1} \, . 
\eeq
If the ALP constitutes all DM, 
the excess reported by the XENON1T experiment implies $f_a / q_e \simeq 10^{10} \GeV$. 
However, it is possible that
the ALP constitute some fraction of DM. 
We denote the fraction by
$r \equiv \Omega_{\rm ALP} / \Omega_{\rm DM}^{\rm (obs)}$
with $\Omega_{\rm ALP}$ being the density parameter of ALP and $\Omega_{\rm DM}^{\rm (obs)}$ ($\simeq 0.24$)  the observed DM abundance. 
Since the XENON1T experiment is sensitive to the combination of $g_{ae}/\sqrt{r}$, 
its excess implies $f_a / q_e \simeq \sqrt{r}\, 10^{10} \GeV$. 
In particular, if, e.g., $r=0.1$, 
we have $f_a / q_e \simeq 3 \times 10^{9} \GeV$.\footnote{If $r=1$ but the DM density is locally suppressed by $10\%$ around the Earth,
 we still get the same $f_a/q_e$.  In this case, the X-ray line signal becomes stronger than the $r= 0.1$ case.}

We are interested in the case where the ALP mass is of order keV, 
 much smaller than the electron mass. 
Thus we can integrate out all the quarks and leptons to investigate the axion coupling to photons in the low energy. 
For the on-shell ALP and photons, 
the relevant term in the effective Lagrangian is given by~\cite{Nakayama:2014cza} (see also \cite{Pospelov:2008jk})
\beq
 \frac{\alpha_{\rm em} m_a^2}{48 \pi f_a} 
 \frac{q_e}{m_e^2}  
 a F_{\mu \nu} \tilde{F}^{\mu \nu} \, . 
 \label{ALP-photon}
\eeq
We can neglect contributions from fermions heavier than electron
because the dominant contribution comes from the lightest one.

\vspace{0.2cm}

\begin{figure}[t!]
\begin{center}  
\includegraphics[width=85mm]{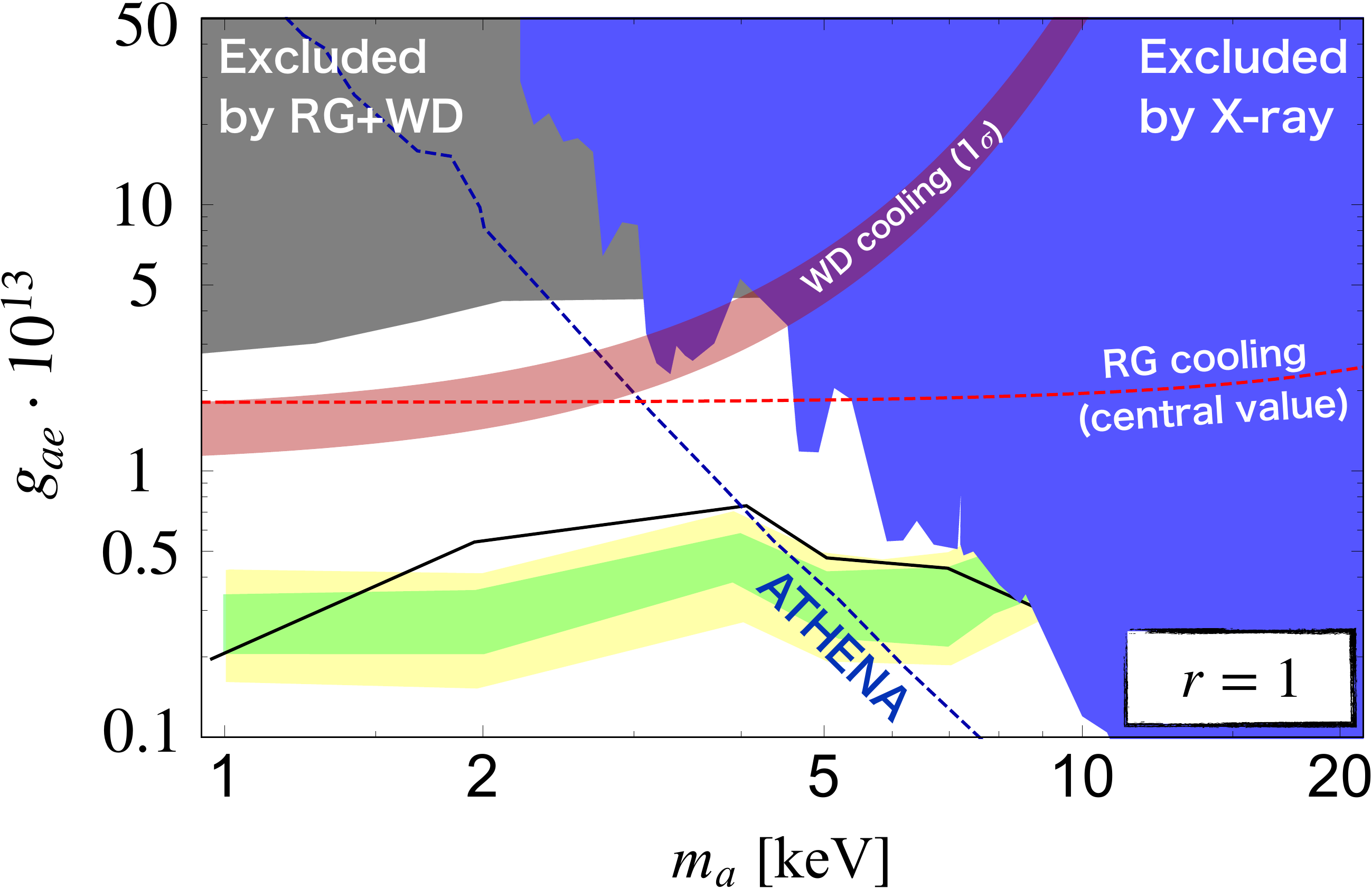}
\\
\includegraphics[width=85mm]{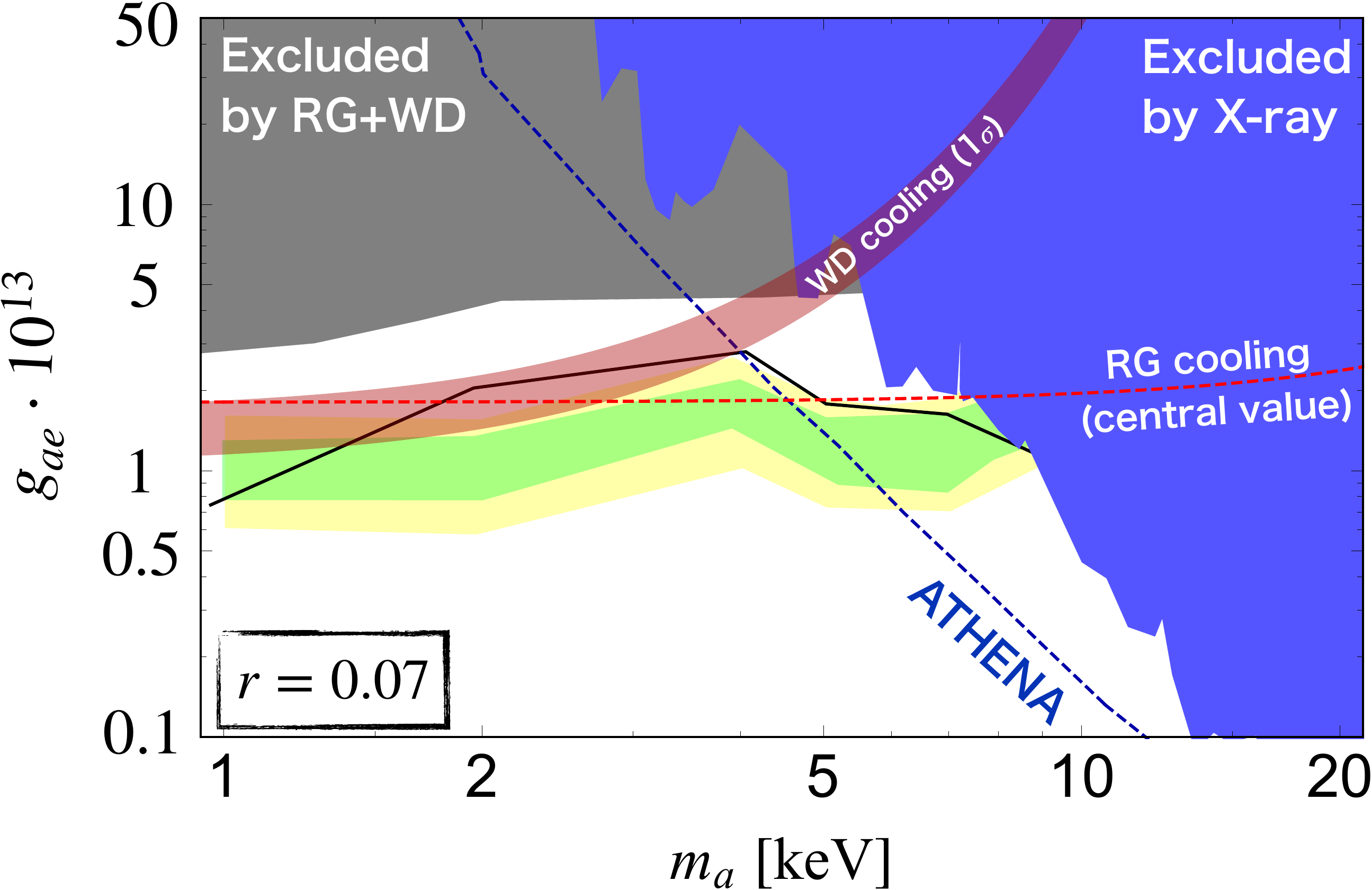}
\end{center}
\caption{
Summary of the present bounds and preferred regions in $g_{ae}$-$m_a$ plane. 
The gray and blue shaded regions are excluded by stellar cooling and X-ray, respectively. 
The future sensitivity by ATHENA is plotted by the blue dashed line~\cite{Neronov:2015kca,Boyarsky:2018tvu}. 
The red shaded region is a preferred region by the WD cooling while the red dashed line is a central value suggested by the RG cooling. 
The XENON1T limits (90\% C.L.) are shown in black with the expected 1 (2) $\sigma$ sensitivities in green (yellow)~\cite{Aprile:2020tmw}. The weaker-than-expected bound around $m_a = (2-3)$\,keV is the region of interest,
where the excess may be explained by the anomaly-free ALP DM.
}
\label{fig:1}
\end{figure}

{\bf ALP decay to photons and X-ray prediction.--} 
The interaction \eq{ALP-photon} leads to the decay of the ALP into two photons. 
The decay rate is calculated as \cite{Nakayama:2014cza}
\beq
 \Gamma_{a \to \gamma \gamma} 
 \simeq 3.5 \times 10^{-57} \, \GeV 
 \lmk \frac{m_a}{2 \KeV} \rmk^7 \lmk \frac{f_a / q_e}{10^{10} \GeV} \rmk^{-2} 
\eeq
There is a constraint on the flux of the X-ray photons produced by the ALP decay. 
It is shown as the blue shaded region in Fig.\,\ref{fig:1}. 
Here, 
we converted the X-ray constraint on the  mixing angle for sterile neutrino DM in \cite{Caputo:2019djj} to $g_{ae}$.
Also shown as the blue dashed line is the combined sensitivity reaches of the Athena X-ray telescope\footnote{https://www.the-athena-x-ray-observatory.eu}~\cite{Neronov:2015kca}, and that obtained by
taking cross correlation of the X-ray line emission 
with galaxy catalogs~\cite{Caputo:2019djj}. The green and yellow bands represent the expected sensitivity of the
XENON1T experiment and the solid black line is the actually obtained limit. The weaker-than-expected limit around
$(2-3)$ keV is the region of interest. Thus, one can see that
the ALP DM hinted from the XENON1T excess is consistent from the current X-ray bound. 
For $m_a=2-3$ keV, the suggested value of the electron coupling $g_{ae}\sim 10^{-13}$
cannot be reached in the future observation. 
If the excess shifts to slightly higher energy in the future experiments (say, $m_a> (4-5)$\,keV) 
with a similar value of $g_{ae}$, the anomaly-free ALP DM scenario may be confirmed by observing the X-ray line emission.

\vspace{0.2cm}

{\bf Stellar cooling anomaly.--}
The evolution of stellar objects, such as WD and RG, 
is modified by the extra cooling via the ALP emission. 
The gray region in Fig.~\ref{fig:1} is 
excluded by such stellar cooling arguments~\cite{Calibbi:2020jvd}.
Interestingly, 
various observations
of WD, RG, and another type of stellar objects can be better fitted in the presence of
excessive cooling, which is known as the stellar cooling anomalies. 
The evolution of WDs is described by a relatively simple and well-known cooling process, 
and the analysis of its luminosity function places relatively strong constraints on the ALP coupling.
We show the WD cooling anomaly reported in Ref.~\cite{Bertolami:2014wua} 
as the red-shaded region in Fig.~\ref{fig:1}.\footnote{
Various uncertainties in the current understanding of the WD evolution are
summarized in Ref.~\cite{Corsico:2019nmr}.
}
On the other hand, the observed brightness of the tip of the RG branch 
in globular clusters also slightly favors a nonzero contribution from the ALP~\cite{Viaux:2013lha}, 
though it is consistent with predictions based on contemporary stellar evolution theory within $1\sigma$. 
In Fig.~\ref{fig:1} we show only the central value of the RG cooling anomaly by the red dashed line.
See the supplemental material for the preferred region of the combination of WD and RG cooling anomalies. 
The XENON1T excess is 
consistent with this bound 
if $r \gtrsim 0.01$.

The regions hinted by the stellar cooling argument are intriguingly close to the excess found in the XENON1T data.
In fact, the overlap becomes even better if 
the fraction of ALP constitutes about 10\% of DM, namely $r\approx 0.1$. This is because the X-ray bound as well as
the XENON1T data scales as $1/\sqrt{r}$ in  Fig.~\ref{fig:1}.
This can be seen in the lower panel in Fig.~\ref{fig:1}, 
where we take $r = 0.07$. 
The red-shaded region and red dashed line are below the black line around $m_a = (2-3) \KeV$. 
Here note that the cooling anomalies require a stronger interaction if the ALP mass is larger than the typical temperature of the stellar system;
the temperature of WD is about $1 \KeV$ 
while that of RG is about $10 \KeV$~\cite{Calibbi:2020jvd}. 
In the supplemental material, we provide  more detailed evaluation of the stellar cooling anomaly and argue that  $r \sim 0.1$ is better than $r = 1$.

Although still under debate~\cite{Hamaguchi:2018oqw}, an excessive cooling was reported in the measurement of neutron star in Cassiopeia A~\cite{Page:2010aw,Shternin:2010qi}. This favors an ALP-nucleon coupling constant of $g_{aN} \approx 4 \times 10^{-10}$~\cite{Leinson:2014ioa}. 
In our model, there is a nonzero ALP-nucleon coupling if the ALP is coupled to quarks
(see e.g. Ref.~\cite{diCortona:2015ldu}). 
Depending on the PQ charge of quarks, 
the coupling can be as large as $m_N/f_a$, where $m_N$ is the neutron mass. Thus we may be able to simultaneously explain the cooling rate of the neutron star as well.

\vspace{0.2cm}

{\bf Production processes.--}
The ALPs can be produced via thermal and non-thermal production processes in the early Universe. 

Relativistic ALPs are produced from 
the scatterings between fermions (quarks and/or leptons) and the Higgs bosons in the thermal bath. 
The resulting ALP abundance is given by~\cite{Nakayama:2014cza} 
\beq
 \Omega_{\rm ALP}^{\rm (th)}  h^2 \sim && 0.01 
 \lmk \frac{T_R}{3 \times 10^5 \GeV} \rmk
 \lmk \frac{m_a}{2 \KeV} \rmk \nonumber\\
 &&\times 
 \lmk \frac{f_a/q_e}{10^{10} \GeV} \rmk^{-2}
 \sum_f \lmk \frac{q_f m_f/q_e}{1 \GeV} \rmk^2 \, , 
 \label{Omega_ALP_th}
\eeq
where $T_R$ is the reheating temperature of the Universe after inflation. 
We define $r^{\rm (th)} \equiv \Omega_{\rm ALP}^{\rm (th)} / \Omega_{\rm DM}^{\rm (obs)}$ for later convenience. 

As we are interested in the case of $m_a = {\cal O}(1) \KeV$, 
thermally produced ALP is warm DM, which has a non-negligible free-streaming velocity at the structure formation.
From the observation of the Lyman-$\alpha$ forest, 
there is a tight constraint on the free-streaming velocity of DM, 
which can be recast into the constraint on the mass of warm DM, and it reads 
$m_a \gtrsim 15 \KeV$ for the case of $r^{\rm (th)} =1$~\cite{Irsic:2017ixq, Kamada:2019kpe}. 
If the fraction of warm component in DM is smaller than unity, 
the constraint becomes weaker, and in fact there is practically no constraint for $r^{\rm (th)} \sim 0.1$ 
(see also Ref.~\cite{Diamanti:2017xfo}).

If the ALP is coupled to the top quark, 
the reheating temperature should be lower than of order $10^2 \GeV$ to satisfy the warm DM bound.\footnote{
For the reheating temperature lower than the top mass,
one has to take account of the entropy production by the inflaton decay.
}
If the ALP is coupled only to electron and muon, on the other hand,
the reheating temperature can be as high as of order $10^8 \GeV$. 
Such  high reheating temperature is favored in light of generating the baryon asymmetry via thermal leptogenesis~\cite{Fukugita:1986hr, Buchmuller:2005eh} or (mild) resonant leptogenesis~\cite{Pilaftsis:1997dr,Buchmuller:1997yu}.

The ALP can be produced also by a non-thermal process, called the misalignment mechanism~\cite{Preskill:1982cy,Abbott:1982af,Dine:1982ah}. 
When the Hubble parameter becomes lower than the ALP mass, 
the ALP starts to oscillate around its potential minimum. 
We denote the temperature at the onset of the ALP oscillation as $T_{\rm osc}$, which is given by 
\beq
 T_{\rm osc} \sim 10^6 \GeV \lmk \frac{m_a}{2 \KeV} \rmk^{1/2} \, .
\eeq
At a temperature above $T_{\rm osc}$, the axion field stays at a field value which is generically deviated from the potential minimum. 
Denoting that the initial oscillation amplitude as $a_{\rm ini} = \theta_* f_a$, 
the oscillation energy of the ALP is given by 
\beq
 \Omega_{\rm ALP}^{\rm (mis)} h^2 \sim 
 && \, 0.1
 \left(\frac{\theta_*}{2}\right)^2 \left(\frac{q_e}{4}\right)^2
 \lmk \frac{f_a/q_e}{10^{10} \GeV} \rmk^{2} 
 \nonumber\\
 && \times
 \left\{ 
 \bea{ll}
 \lmk \frac{T_R}{10^6 \GeV} \rmk \quad &{\rm for}~~ T_R \lesssim T_{\rm osc}
 \vspace{0.1cm}\\
 \lmk \frac{m_a}{2 \KeV} \rmk^{1/2} 
 \quad &{\rm for}~~T_R \gtrsim T_{\rm osc}
 \eea
 \right. \, .
\eeq
The resulting relic abundance depends on $q_e$ once we choose $f_a / q_e$ to be the value favored by the XENON1T result. 
Thus, the ALP can explain the observed amount of DM with $T_R>T_{\rm osc}$ for $q_e$ and $\theta_*$ of order unity.
Due to the warm DM limit, $T_{R}$ may not be higher than $10^6$ GeV, depending on the coupling with other SM particles.
In this case, one may take $q_e$ much larger than unity to realize $ \Omega_{\rm ALP}^{\rm (mis)} \sim 0.1$
by using, e.g., the clockwork mechanism~\cite{Choi:2015fiu,Kaplan:2015fuy,Giudice:2016yja},
while $f_a/q_e$ is fixed.

We note that $r \simeq 0.1$ is interesting in light of cooling anomaly. 
In this case, we do not need to produce ALPs by the misalignment mechanism 
because $ r =r^{\rm (th)} \sim 0.1$ satisfies the Lyman-$\alpha$ constraint.

\vspace{0.2cm}

{\bf Isocurvature constraint.--}
We implicitly assume that the PQ symmetry is spontaneously broken before inflation. 
In this case, the ALP predicts an isocurvature perturbation due to quantum fluctuations during inflation. 
The fluctuation is proportional to the Hubble parameter of inflation, $H_{\rm inf}$, so that the constraint on the isocurvature perturbation can be rewritten as~\cite{Akrami:2018odb}
\beq
 H_{\rm inf} \lesssim 3 \times 10^5\theta_* r^{-1} \left( \frac{f_a}{10^{10}\GeV} \right){\rm\, GeV} \, .
\eeq

Note that the Universe must be reheated after inflation and the energy of the thermal bath cannot exceed the energy scale of inflation. Thus the upper bound on $H_{\rm inf}$ implies that the reheating temperature must satisfy 
\beq
 T_R \lesssim 8\times 10^{11} {\, \rm GeV} \sqrt{\frac{H_{\rm inf}}{10^6 {\,\rm GeV}}} \, .
\eeq
in the case of the instantaneous reheating. This is consistent with the scenario of ALP production explained above. 
A relatively low reheating temperature ($T_R \ll f_a$) is also consistent with 
the assumption that the PQ symmetry is spontaneously broken before inflation and is never restored after inflation.

\vspace{0.2cm}

{\bf Discussion.--}
As we discussed above, 
the XENON1T excess favors a larger value of $g_{ae} \approx 10^{-13}$
when the ALP constitutes only about 10\% of DM. 
In this case, all ALPs can be thermally produced. Although it can evade the bound from the Lyman-$\alpha$ forest data, 
we expect that there is still some effect on the small-scale structure formation that can be searched for by future observations.

We note that 
it might be natural to consider 
the ALP as a subdominant component of DM in light of the anthropic principle~\cite{Davies, Barrow, Barrow:1988yia, Hawking:1987en, Weinberg:1987dv} and the string axiverse~\cite{Green:1987sp,Svrcek:2006yi}. 
According to the string axiverse, 
there are many ALPs with logarithmically distributed masses in the low-energy effective theory, 
one of which may be identified as the very ALP considered in this letter. 
We expect that there are many other ALPs, 
some of which may be very light and behave as dark radiation~\cite{Takahashi:2019ypv},
while some others may be heavy and contribute to DM. 
The relic abundance of the latter ALPs 
may depend on parameters that vary in the multiverse. 
According to the selection effect due to the anthropic condition, 
the abundance of these ALPs (and the other DM candidates) should not be much larger than the observed amount of DM to avoid the overproduction of cosmological structures in 
the habitable universes (see e.g. Ref.~\cite{Tegmark:2004qd}). 
In such multi-component DM scenario, where there are several DM candidates with energy densities smaller than the observed one,
it may be natural that our ALP constitutes only ${\cal O}(10)\%$ fraction of DM.
The reason that the ALP decay constant is close to the stellar cooling bound may also be explained by the anthropic principle. 

In a class of ALP inflation models~\cite{Daido:2017wwb, Daido:2017tbr, Takahashi:2019qmh}, an ALP plays both roles of DM and inflaton. Interestingly, the mass is predicted as $m_a\sim 1 \KeV q_e^2 (10^{-13}/g_{ae})^{2}$ to fit the cosmic microwave background data if the ALP couples to electrons. The reheating can be successful if the inflaton is also coupled to heavy fermions. 
If the ALP is thermalized after reheating, its abundance must be diluted by the entropy production.
Alternatively, if the reheating is incomplete as in the original ALP miracle scenario~\cite{Daido:2017wwb, Daido:2017tbr}, some amount of coherent oscillations 
remain and explain all DM. This possibility will be discussed elsewhere.

Although we take $m_a = 2\, \text{-} \, 3 \KeV$ as a reference value in this letter, we note that a heavier ALP is well fitted by our model. 
It is still possible that improved data by XENONnT will favor a heavier ALP mass. 
If the ALP mass is $(4-5)$ keV or heavier, 
we expect that the X-ray signals from the ALP decay will be observed by ATHENA.

%
\begin{acknowledgments}
We thank Jure Zupan and Andrea Caputo for useful comments.
F.~T. was supported by JSPS KAKENHI Grant Numbers
17H02878, 20H01894 and by World Premier International Research Center Initiative (WPI Initiative), MEXT, Japan. 
M.~Y. was supported by Leading Initiative for Excellent Young Researchers, MEXT, Japan. W.~Y was supported by JSPS KAKENHI Grant No. 16H06490.
\end{acknowledgments}
%

\vspace{1cm}

\bibliography{references}

\end{document}



\newcommand{\vev}[1]{ \left\langle {#1} \right\rangle }
\newcommand{\eV}{ \ {\rm eV} }
\newcommand{\KeV}{ \ {\rm keV} }
\newcommand{\MeV}{\  {\rm MeV} }
\newcommand{\GeV}{\  {\rm GeV} }
\newcommand{\TeV}{\  {\rm TeV} }
\newcommand{\1}{\mbox{1}\hspace{-0.25em}\mbox{l}}

\newcommand{\lmk}{\left(}  
\newcommand{\rmk}{\right)}
\newcommand{\lkk}{\left[}  
\newcommand{\rkk}{\right]}
\newcommand{\lhk}{\left \{ }  
\newcommand{\rhk}{\right \} }
\newcommand{\del}{\partial}  
\newcommand{\la}{\left\langle} 
\newcommand{\ra}{\right\rangle}
\newcommand{\half}{\frac{1}{2}}

\newcommand{\bea}{\begin{array}}
\newcommand{\eea}{\end{array}}
\newcommand{\beq}{\begin{eqnarray}}
\newcommand{\eeq}{\end{eqnarray}}
\newcommand{\eq}[1]{Eq.~(\ref{#1})}

\newcommand{\dd}{\mathrm{d}}
\newcommand{\Mpl}{M_{\rm Pl}}
\newcommand{\mg}{m_{3/2}}
\newcommand{\abs}[1]{\left\vert {#1} \right\vert}
\newcommand{\mphi}{m_{\phi}}
\newcommand{\Hz}{\ {\rm Hz}}
\newcommand{\for}{\quad \text{for }}
\newcommand{\Min}{\text{Min}}
\newcommand{\Max}{\text{Max}}
\newcommand{\Kahler}{K\"{a}hler }
\newcommand{\cphi}{\varphi}
\newcommand{\Tr}{\text{Tr}}
\newcommand{\diag}{{\rm diag}}

\def\lrf#1#2{ \left(\frac{#1}{#2}\right)}
\def\lrfp#1#2#3{ \left(\frac{#1}{#2} \right)^{#3}}
\def\lrp#1#2{\left( #1 \right)^{#2}}
\def\REF#1{Ref.~\cite{#1}}
\def\SEC#1{Sec.~\ref{#1}}
\def\FIG#1{Fig.~\ref{#1}}
\def\EQ#1{Eq.~(\ref{#1})}
\def\EQS#1{Eqs.~(\ref{#1})}
\def\TEV#1{10^{#1}{\rm\,TeV}}
\def\GEV#1{10^{#1}{\rm\,GeV}}
\def\MEV#1{10^{#1}{\rm\,MeV}}
\def\KEV#1{10^{#1}{\rm\,keV}}
\def\blue#1{\textcolor{blue}{#1}}
\def\red#1{\textcolor{red}{#1}}

\title{Supplemental material: Models of anomaly-free ALP}




\maketitle
  
  \makeatletter
    \renewcommand{\theequation}{%
    \thesection.\arabic{equation}}
    \@addtoreset{equation}{section}
  \makeatother

 \makeatletter
    \renewcommand{\thefigure}{%
    \thesection.\arabic{figure}}
    \@addtoreset{figure}{section}
  \makeatother

Throughout the \textit{Letter} we study the ALP DM as a potential source for the XENON1T excess, and discuss
its implications for the stellar cooling anomaly, assuming that 
the PQ symmetry is free from the U(1)$_{\rm PQ}$-U(1)$_{\rm em}$-U(1)$_{\rm em}$ anomaly. 
In this supplemental material, we revisit the leptophilic ALP DM model~\cite{Nakayama:2014cza} as well as a model
based on a two Higgs doublet model (2HDM) as concrete examples of anomaly-free ALPs. 
We also provide detailed evaluation of the stellar cooling anomaly 
and its comparison  with the XENON1T excess for various values of the fraction of ALP DM.

\section{Models of anomaly-free ALP}

First, we briefly review a model of Ref.~\cite{Nakayama:2014cza} 
where the ALP is coupled only to electron and muon. 
The charge assignment is $Q(e_i) = (-2, 2, 0)$ and $Q(l_i) = (1,-1,0)$, 
where $i$ represents the generation of quarks and leptons. 
We also introduce three Higgs doublets, $H(0)$ and $H(\pm 3)$. 
One can easily check that there is no U(1)$_{\rm PQ}$-U(1)$_{\rm em}$-U(1)$_{\rm em}$ anomaly in this model. 
The charged lepton Yukawa interactions are given by 
\beq
 y_1 \bar{e}_R l_1 H(-3) 
 +
 y_2 \bar{\mu}_R l_2 H(3) 
 + 
 y_3 \bar{\tau}_R l_3 H(0) + {\rm h.c.}
\eeq
One can also assign the flavor charge on the right-handed neutrinos as $Q(N_i) = (1,-1,0)$,
and introduce three B-L Higgs fields $\phi(0),\phi(1)$, and $\phi(2)$
to generate the neutrino mass 
through the seesaw mechanism \cite{Yanagida:1979as,Yanagida:1980xy,GellMann:1980vs,Minkowski:1977sc}. 
The PQ symmetry is broken by the B-L Higgs fields whose vacuum expectation values are assumed to be
of the order of $10^{9-10}$\,GeV. The ALP resides mainly in the phases of the B-L Higgs fields.

Here we comment on the possible origin of the mass of the ALP.
The ALP mass arises from an explicit breaking of the PQ symmetry. 
Noting that the keV scale comes from two different energy scales of $10^{10}\GeV$ and the electroweak scale, 
we consider the following PQ-breaking term: 
\beq
 m_H^2 H(-3)^\dagger H(3) + {\rm h.c.}
\eeq
This leads to $m_a = {\cal O}(1) \KeV$ if the size of the expectation value of the above operator is
of order [the weak scale]$^4$ and $f_a \sim 10^{10} \GeV$. This is realized 
for a certain choices of $m_H$, $ \la H(-3) \ra$, and $\la H(3) \ra $. 
It is interesting to note that the keV scale favored by the XENON1T anomaly 
comes from the PQ breaking scale and the electroweak scale without additional small parameters. 
We assume that the Higgs fields other than the lightest one 
are heavy enough to evade the collider constraints, but some of them may be within the reach of future collider experiments.

There are a variety of models of the anomaly free ALP 
coupled to other quarks and leptons. 
For example, 
one can simply assume that the fermions are universally charged under the PQ symmetry. 
The ALP mass is generated by introducing higher dimensional terms which explicitly break the PQ symmetry.
This is the case in a two Higgs doublet model with Higgs $H_u$ and $H_d$ 
to be both charged under the PQ symmetry. The mixing term, ${\cal L}\supset H_u H_d$, is not invariant under the PQ symmetry and thus is obtained through the spontaneous breaking of the PQ symmetry.  Their VEVs are given as $\vev{H_{u}}=v \sin\beta$, and $\vev{H_{u}}=v \cos\beta$ with $v\approx 174\,$GeV. 
In the low-energy effective theory after integrating out heavy Higgs and PQ breaking fields, one obtains
\beq
     H_u & = \tilde{H} \sin\beta  \exp{\left[i Q[{H_u}] \frac{a}{f_a}\right]},\\
    H_d & = H \cos\beta  \exp{\left[i Q[{H_d}] \frac{a}{f_a}\right]} \, .
\eeq
where $Q[H_d]/Q[H_u] = \tan^2 \beta$. 
Here $H$ and $\tilde{H}$ both represent the SM Higgs doublet with $\tilde{H} = i \sigma_2 H^*$. 
If either $H_u$ or $H_d$ does not couple to the SM fermions, {\it a la} type I (fermiophobic) 2HDM, the PQ charges of the SM fermions are 
universal~\cite{Takahashi:2019qmh}. 
Then one can easily check that the anomaly is automatically cancelled in this model.

Since the ALP couples to all the SM fermions in this model, 
the thermal production is very efficient, and
one needs to have a relatively low reheating temperature to suppress the thermal production.
Specifically, one needs $T_R\lesssim 100{\rm GeV}$ to have  $r \lesssim 0.1$. 
For $T_R<100\,$GeV, on the other hand, the misalignment contribution is suppressed due to entropy production from the inflaton decay (i.e., $T_R < T_{\rm osc}$). 
Therefore, the ALP cannot explain all the DM. Still, it is possible to explain the XENON1T excess if the warm component is significant, $r \simeq r^{\rm (th)} \sim 0.1$.
In such a low reheating temperature, the baryon asymmetry of the Universe may be generated by, e.g., 
the Affleck-Dine mechanism~\cite{Affleck:1984fy,Dine:1995kz} 
or the spontaneous baryogenesis around the electroweak phase transition~\cite{Servant:2014bla,Jeong:2018jqe,Co:2019wyp,Croon:2019ugf,Domcke:2020kcp,Co:2020xlh}.

Notice that the model contains not only the ALP-electron coupling but also the ALP-nucleon coupling
\beq
g_{aN}\sim 0.5 \frac{m_N}{m_e}g_{ae}\sim 2\times 10^{-10}\left(\frac{g_{ae}}{2\times 10^{-13}}\right).
\eeq
This is around the aforementioned value hinted by the possible cooling anomaly of the neutron star in Cassiopeia A~\cite{Leinson:2014ioa}.

\section{Stellar cooling anomalies and the fraction of ALP DM}

In this \textit{Letter}, we argue that the stellar cooling anomaly, as well as the excess of XENON1T, can be better 
explained if the anomaly-free ALP DM constitutes about 10\% of the DM.
Here we provide a more detailed assessment of the stellar cooling anomaly to discuss its implications
for the fraction of ALP DM.

\begin{figure}[t!]
\begin{center}  
\includegraphics[width=85mm]{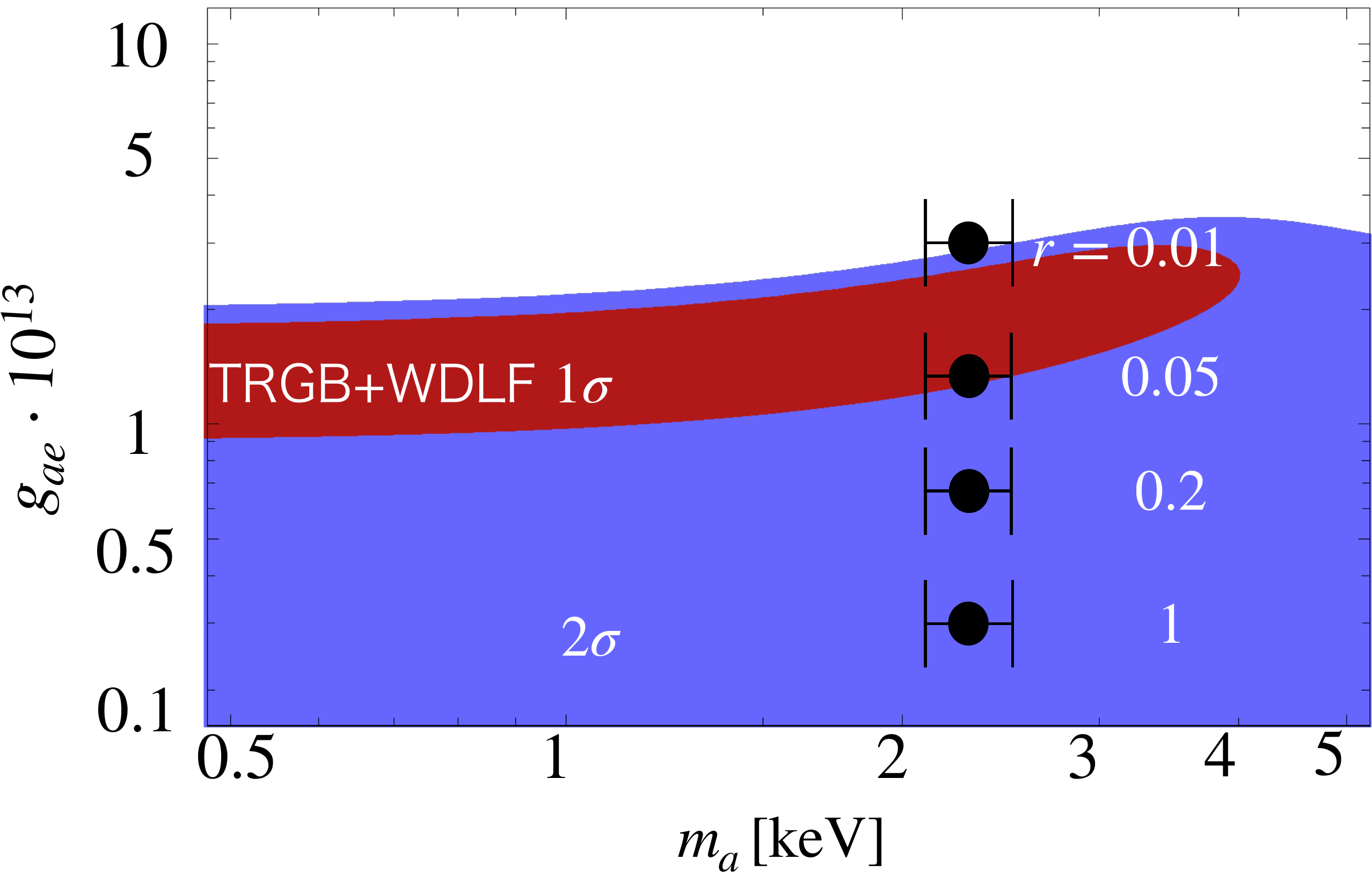}
\end{center}
\caption{
The ALP mass and coupling favored by the combined analysis of the WD and RG cooling hints. 
The red (blue) region is favored at the $1\sigma$ ($2\sigma$) level. 
The mass and coupling suggested by the XENON1T excess  are shown by the black dots for
 $r=0.01,0.05,0.2,1$ from top to bottom.  
}
\label{fig:sup}
\end{figure}

We estimate the $\chi^2$ distribution by combining the cooling hints from the WD luminosity function, $10^{26}g_{ae}^2/4\pi=0.156 \pm  0.068 $,  
and from the tip of RG branch, $10^{26}g_{ae}^2/(4\pi)=0.26\pm 0.28$.
The red band in Fig.\,\ref{fig:sup} shows the $1\sigma$  favored region, and
the blue region is  the $2\sigma$ allowed region which extends down to $g_{ae}=0$.
 Also shown are the ALP mass and coupling suggested by the XENON1T excess, $m_a = 2.3 \pm 0.2$\,keV and $g_{ae}\approx 3 \times 10^{-14}/\sqrt{r}$ 
 (see v2 of Ref.\,~\cite{Aprile:2020tmw}), for various values of $r$. We take $r=0.01,0.05,0.2,1$ from top to bottom (black dots). 
One can see that the agreement between the stellar cooling anomalies and the XENON1T excess
becomes better for $r=0.01-0.1$.

\vspace{1cm}

\bibliography{references}